\input harvmac   
\noblackbox   


\def\p{\partial}

\def\CN{{\cal N}}

\def\CL{{\cal L}}

\def\CS{{\cal S}}

\def\CS{{\cal S }}

\def\G{\Gamma}    


\font\manual=manfnt     
\def\dbend{\lower3.5pt\hbox{\manual\char127}}

\def\IZ{\relax\ifmmode\mathchoice    
{\hbox{\cmss Z\kern-.4em Z}}{\hbox{\cmss Z\kern-.4em Z}}    
{\lower.9pt\hbox{\cmsss Z\kern-.4em Z}} {\lower1.2pt\hbox{\cmsss    
Z\kern-.4em Z}}\else{\cmss Z\kern-.4em Z}\fi}    
\def\half {{1\over 2}}

\def\p{\partial}    
\def\pb{\bar{\partial}}    
\def\bar{\overline}    
\def\CS{{\cal S}}    
\def\CN{{\cal N}}    
    
\def\rt2{\sqrt{2}}    
\def\irt2{{1\over\sqrt{2}}}

\def\t{\tilde}    
\def\ndt{\noindent}    
    
\def\b{\beta}    
\def\a{\alpha}

\font\cmss=cmss10    
\font\cmsss=cmss10 at 7pt    
\def\IL{\relax{\rm I\kern-.18em L}}    
\def\IH{\relax{\rm I\kern-.18em H}}    
\def\IR{\relax{\rm I\kern-.18em R}}    
\def\inbar{\vrule height1.5ex width.4pt depth0pt}    
\def\IC{\relax\hbox{$\inbar\kern-.3em{\rm C}$}}    
\def\rlx{\relax\leavevmode}    
\def\ZZ{\rlx\leavevmode\ifmmode\mathchoice{\hbox{\cmss Z\kern-.4em Z}}    
 {\hbox{\cmss Z\kern-.4em Z}}{\lower.9pt\hbox{\cmsss Z\kern-.36em Z}}    
 {\lower1.2pt\hbox{\cmsss Z\kern-.36em Z}}\else{\cmss Z\kern-.4em    
 Z}\fi}     
\def\IZ{\relax\ifmmode\mathchoice    
{\hbox{\cmss Z\kern-.4em Z}}{\hbox{\cmss Z\kern-.4em Z}}    
{\lower.9pt\hbox{\cmsss Z\kern-.4em Z}}    
{\lower1.2pt\hbox{\cmsss Z\kern-.4em Z}}\else{\cmss Z\kern-.4em    
Z}\fi}    
    

\def\G{\Gamma}

\font\manual=manfnt     
\def\dbend{\lower3.5pt\hbox{\manual\char127}}

\def\IZ{\relax\ifmmode\mathchoice    
{\hbox{\cmss Z\kern-.4em Z}}{\hbox{\cmss Z\kern-.4em Z}}    
{\lower.9pt\hbox{\cmsss Z\kern-.4em Z}} {\lower1.2pt\hbox{\cmsss    
Z\kern-.4em Z}}\else{\cmss Z\kern-.4em Z}\fi}    
\def\half {{1\over 2}}

\def\pb{\bar{\partial}}    
\def\bar{\overline}    

\def\rt2{\sqrt{2}}    
\def\irt2{{1\over\sqrt{2}}}

\def\t{\tilde}    
\def\T{\widetilde}
\def\ndt{\noindent}

\let\includefigures=\iftrue
\let\useblackboard=\iftrue
\newfam\black

\includefigures
\message{If you do not have epsf.tex (to include figures),}
\message{change the option at the top of the tex file.}
\input epsf
\def\figin{\epsfcheck\figin}\def\figins{\epsfcheck\figins}
\def\epsfcheck{\ifx\epsfbox\UnDeFiNeD
\message{(NO epsf.tex, FIGURES WILL BE IGNORED)}
\gdef\figin##1{\vskip2in}\gdef\figins##1{\hskip.5in}
\else\message{(FIGURES WILL BE INCLUDED)}%
\gdef\figin##1{##1}\gdef\figins##1{##1}\fi}
\def\DefWarn#1{}
\def\figinsert{\goodbreak\midinsert}
\def\ifig#1#2#3{\DefWarn#1\xdef#1{fig.~\the\figno}
\writedef{#1\leftbracket fig.\noexpand~\the\figno}%
\figinsert\figin{\centerline{#3}}\medskip\centerline{\vbox{
\baselineskip12pt\advance\hsize by -1truein
\noindent\footnotefont{\bf Fig.~\the\figno:} #2}}
\bigskip\endinsert\global\advance\figno by1}
\else
\def\ifig#1#2#3{\xdef#1{fig.~\the\figno}
\writedef{#1\leftbracket fig.\noexpand~\the\figno}%
\global\advance\figno by1}
\fi

\def\doublefig#1#2#3#4{\DefWarn#1\xdef#1{fig.~\the\figno}
\writedef{#1\leftbracket fig.\noexpand~\the\figno}%
\figinsert\figin{\centerline{#3\hskip1.0cm#4}}\medskip\centerline{\vbox{
\baselineskip12pt\advance\hsize by -1truein
\noindent\footnotefont{\bf Fig.~\the\figno:} #2}}
\bigskip\endinsert\global\advance\figno by1}



\lref\sameer{
S.~Murthy,
``Notes on non-critical superstrings in various dimensions,''
JHEP {\bf 0311}, 056 (2003)
[arXiv:hep-th/0305197].
}

\lref\gavanar{
  E.~Gava, K.~S.~Narain and M.~H.~Sarmadi,
  ``Little string theories in heterotic backgrounds,''
  Nucl.\ Phys.\ B {\bf 626}, 3 (2002)
  [arXiv:hep-th/0112200].
}

\lref\larsei{
  J.~L.~Davis, F.~Larsen and N.~Seiberg,
  ``Heterotic strings in two dimensions and new stringy phase transitions,''
  JHEP {\bf 0508}, 035 (2005)
  [arXiv:hep-th/0505081].
}

\lref\SeibergNK{
  N.~Seiberg,
  ``Long strings, anomaly cancellation, phase transitions, T-duality and
  locality in the 2d heterotic string,''
  JHEP {\bf 0601}, 057 (2006)
  [arXiv:hep-th/0511220].
}

\lref\DavisQI{
  J.~L.~Davis,
  ``The moduli space and phase structure of heterotic strings in two
  dimensions,''
  arXiv:hep-th/0511298.
}

\lref\kapustin{
  M.~Gremm and A.~Kapustin,
  ``Heterotic little string theories and holography,''
  JHEP {\bf 9911}, 018 (1999)
  [arXiv:hep-th/9907210].
}

\lref\kutsei{
  D.~Kutasov and N.~Seiberg,
  ``Noncritical Superstrings,''
  Phys.\ Lett.\ B {\bf 251}, 67 (1990).
}

\lref\kutgiv{
  A.~Giveon and D.~Kutasov,
  ``Little string theory in a double scaling limit,''
  JHEP {\bf 9910}, 034 (1999)
  [arXiv:hep-th/9909110] 
\semi 
 A.~Giveon, D.~Kutasov and O.~Pelc,
  ``Holography for non-critical superstrings,''
  JHEP {\bf 9910}, 035 (1999)
  [arXiv:hep-th/9907178].
}

\lref\KutasovPV{
  D.~Kutasov,
  ``Some properties of (non)critical strings,''
  arXiv:hep-th/9110041.
}

\lref\narain{
  K.~S.~Narain,
  ``New Heterotic String Theories In Uncompactified Dimensions $<$ 10,''
  Phys.\ Lett.\ B {\bf 169}, 41 (1986).
}

\lref\itzhaki{
  N.~Itzhaki, D.~Kutasov and N.~Seiberg,
  ``Non-supersymmetric deformations of non-critical superstrings,''
  arXiv:hep-th/0510087.
}

\lref\lust{
  D.~Lust and S.~Theisen,
  ``Lectures On String Theory,''
  Lect.\ Notes Phys.\  {\bf 346}, 1 (1989).
}

\lref\eguchiI{
  T.~Eguchi and Y.~Sugawara,
  ``Conifold type singularities, N = 2 Liouville and SL(2:R)/U(1) theories,''
  JHEP {\bf 0501}, 027 (2005)
  [arXiv:hep-th/0411041]
\semi
  T.~Eguchi and Y.~Sugawara,
  ``Modular invariance in superstring on Calabi-Yau n-fold with A-D-E
  singularity,''
  Nucl.\ Phys.\ B {\bf 577}, 3 (2000)
  [arXiv:hep-th/0002100].
}

\lref\SeibergBX{
  N.~Seiberg,
  ``Observations on the moduli space of two dimensional string theory,''
  JHEP {\bf 0503}, 010 (2005)
  [arXiv:hep-th/0502156].
}

\lref\SeibergEB{
  N.~Seiberg,
  ``Notes On Quantum Liouville Theory And Quantum Gravity,''
  Prog.\ Theor.\ Phys.\ Suppl.\  {\bf 102}, 319 (1990).
}

\lref\EguchiIK{
  T.~Eguchi and Y.~Sugawara,
  ``Conifold type singularities, N = 2 Liouville and SL(2,R)/U(1) theories,''
  JHEP {\bf 0501}, 027 (2005)
  [arXiv:hep-th/0411041].
}

\lref\deAlwisPR{
  S.~P.~de Alwis, J.~Polchinski and R.~Schimmrigk,
  ``Heterotic Strings With Tree Level Cosmological Constant,''
  Phys.\ Lett.\ B {\bf 218}, 449 (1989).
}

\lref\McGuiganQP{
  M.~D.~McGuigan, C.~R.~Nappi and S.~A.~Yost,
  ``Charged black holes in two-dimensional string theory,''
  Nucl.\ Phys.\ B {\bf 375}, 421 (1992)
  [arXiv:hep-th/9111038].
}

\lref\GiveonHM{
  A.~Giveon, E.~Rabinovici and A.~A.~Tseytlin,
  ``Heterotic string solutions and coset conformal field theories,''
  Nucl.\ Phys.\ B {\bf 409}, 339 (1993)
  [arXiv:hep-th/9304155].
}

\lref\GiveonZZ{
  A.~Giveon and A.~Sever,
  ``Strings in a 2-d extremal black hole,''
  JHEP {\bf 0502}, 065 (2005)
  [arXiv:hep-th/0412294].
}

\lref\HellermanZM{
  S.~Hellerman,
  ``On the landscape of superstring theory in D $>$ 10,''
  arXiv:hep-th/0405041.
}

\lref\fzz{V.~A.~Fateev, A.~B.~Zamolodchikov and Al.~B.~Zamolodchikov, Unpublished notes.}

\lref\kutgivads{A.~Giveon and D.~Kutasov, ``Notes on $AdS_3$'', Nucl.\ Phys.\ B {\bf 621}, 303 (2002)  [arXiv:hep-th/9909110]. }

\lref\wittenbh{E.~Witten, ``String theory and black holes'', Phys.\ Rev.\ D {\bf 44}, 314 (1991) \semi 
G.~Mandal, A.~M.~Sengupta and S.~R.~Wadia, ``Classical solutions of two-dimensional string theory'', Mod.\ Phys.\ Lett. A {\bf 6}, 1685 (1991) \semi
S.~Elitzur, A.~Forge and E.~Rabinovici, ``Some global aspects of string compactifications'', Mod.\ Phys.\ Lett. A {\bf 6}, 1685 (1991).}

\lref\dvv{R.~Dijkgraaf, H.~Verlinde and E.~Verlinde, ``String propagation in a black hole geometry'',  Nucl.\ Phys.\ B {\bf 371}, 269 (1992).}

\lref\eguchi{T.~Eguchi and Y.Sugawara, ``Modular invariance in superstring on Calabi-Yau n-fold with A-D-E singularity'',  Nucl.\ Phys.\ B {\bf 577}, 3 (2000) [arXiv:hep-th/0002100].  }

\lref\mizo{S.~Mizoguchi, ``Modular invariant critical superstrings on four-dimensional Minkowski space $\times$ two-dimensional black hole'', JHEP {\bf 0004} 14 (2000) [arXiv:hep-th/0003053]. }

\lref\bilal{A.~Bilal and J.~L.~Gervais, ``New critical dimensions for string theories'', Nucl.\ Phys.\ B {\bf 284}, 397 (1987). }

\lref\farkas{H.~M.~Farkas and I.~Kra, ``Theta constants, Riemann surfaces and the modular group'', Graduate studies in mathematics, Vol.37. Amer.\ Math.\ Soc.\ }

\lref\horikap{K.~Hori and A.~Kapustin, ``Duality of the fermionic 2-d black hole and $\CN=2$ Liouville theory as mirror symmetry'', JHEP {\bf 0108} 045 (2001) [arXiv:hep-th/0104202].}

\lref\oogvafa{H.~Ooguri and C.~Vafa, ``Two-dimensional black hole and singularities of CY manifolds'', Nucl.\ Phys.\ B {\bf 463}, 55 (1996) [arXiv:hep-th/9511164]. }

\lref\dilholo{O.~Aharony, M.~Berkooz, D.~Kutasov and N.~Seiberg, ``Linear dilatons, NS5-branes and holography'', JHEP {\bf 9810} 004 (1998) [arXiv:hep-th/9808149]. }

\lref\seiberg{N.~Seiberg, ``Notes on quantum Liouville theory and quantum gravity'', Prog.\ Theor.\ Phys.\ Suppl {\bf 102} 319 (1990).}

\lref\maldoog{J.~Maldacena and H.~Ooguri,``Strings in $AdS_3$ and $SL_2(\IR)$ WZW model 1: The spectrum'', J.\ Math.\ Phys.\ {\bf 42} 2929 (2001)  [arXiv:hep-th/0001053]. }

\lref\fms{D.~Friedan, E.~Martinec and S.~Shenker, ``Conformal invariance, supersymmetry and string theory'', Nucl.\ Phys.\ B {\bf 271}, 93 (1986). }

\lref\chs{C.~Callan, J.~Harvey and A.~Strominger, ``Supersymmetric string solitons'', [arXiv:hep-th/9112030], in Trieste 1991, Proceedings, String theory and Quantum Gravity 1991, 208.}

\lref\polchinski{J.~Polchinski, ``String theory, Vol. I, II'', Cambridge University Press (1998).}

\lref\kleb{C.~P.~Herzog, I.~R.~Klebanov and P.~Ouyang, ``Remarks on the warped deformed conifold'',  [arXiv:hep-th/0108101].}

\lref\seiwitt{N.~Seiberg and E.~Witten, ``The $D1/D5$ system and singular CFT'', JHEP {\bf 9904} 017 (1999), [arxiv:hepth/9903224].}

\lref\mukhi{K.~Dasgupta and S.~Mukhi,``Brane constructions, conifolds and M-theory'', Nucl.\ Phys.\ B {\bf 551}, 204 (1999), [arxiv:hepth/9811139]. }

\lref\tong{D.~Tong, ``NS5-branes, T-duality and worldsheet instantons'', JHEP {\bf 0207} 013 (2002) [arXiv:hep-th/0204186].}

\lref\kazsuz{Y.~Kazama and H.~Suzuki, ``New $N=2$ superconformal field theories and superstring compactification'', Nucl.\ Phys.\ B {\bf 321}, 232 (1989). }

\lref\harmoo{R.~Gregory, J.~Harvey and G.~Moore,``Unwinding strings and T-duality of Kaluza-Klein and H-monopoles'', Adv.\ Theor.\ Math.\ Phys.\ {\bf 1} 283 (1997), [arXiv:hep-th/9708086]. }

\lref\klemvafa{A.~Klemm, W.~Lerche, P.~Mayr, C.~Vafa and N.~Warner, ``Self-dual strings and $N=2$ supersymmetric field theory'', Nucl.\ Phys.\ B {\bf 477}, 746 (1996),  [arXiv:hep-th/9604034]\semi
S.~Gukov, C.~Vafa and E.~Witten, ``CFT's from Calabi-Yau four-folds'', Nucl.\ Phys.\ B {\bf 584}, 69 (2000),  [arXiv:hep-th/9906070]. }

\lref\polya{A.~M.~Polyakov, ``Quantum Geometry of Bosonic Strings'', Phys.\ Lett.\ B {\bf 103}, 207 (1981).}

\lref\BershadskyZS{
M.~Bershadsky and I.~R.~Klebanov,
``Partition functions and physical states in two-dimensional quantum gravity and supergravity,''
Nucl.\ Phys.\ B {\bf 360}, 559 (1991).}

\lref\TongIK{
D.~Tong,
``Mirror mirror on the wall: On two-dimensional black holes and Liouville  theory,''
JHEP {\bf 0304}, 031 (2003)
[arXiv:hep-th/0303151].
}


\Title{\vbox{\baselineskip12pt\hbox{hep-th/0603121}\hbox{IC/2006/008}
}} 
{\vbox{\centerline{Non-critical Heterotic superstrings in Various Dimensions.} 
}}

{\vskip -20pt\baselineskip 14pt   
\centerline{
Sameer Murthy$^{1}$ \footnote{}{$^1$\tt smurthy@ictp.it}  
}  

\bigskip  

\centerline{\sl Abdus Salam International Center for Theoretical Physics}  
\centerline{\sl Strada Costiera 11, Trieste, 34014, Italy.}  
\bigskip\bigskip

\centerline{\bf Abstract}
We construct heterotic string theories on spacetimes of the form $R^{d-1,1} \times \CN=2$ linear dilaton, where $d=6,4,2,0$.  There are two lines of supersymmetric theories descending from the two supersymmetric ten-dimensional heterotic theories. These have gauge groups which are lower rank  subgroups  of  $E_{8} \times E_{8}$ and $SO(32)$. On turning on a $(2,2)$ deformation which makes the two dimensional part a smooth $SL_{2}(\IR)/U(1)$ supercoset, the gauge groups get broken further. In the deformed theories, there are non-trivial moduli which are charged under the surviving gauge group in the case of $d=6$. We construct the marginal  operators on the worldsheet corresponding to these moduli. 
\noindent   
}

\Date{March 2006} 

\newsec{Introduction and summary}

Non-critical superstrings  in various dimensions $\IR^{d-1,1}$ were constructed in \kutsei\ as a theory using in addition to the flat space coordinates, the $\CN=2$ Liouville theory on the worldsheet. Since then, the type II theories have been studied in detail from various points of view \refs{\kutgiv, \sameer, \mizo,  \eguchi}. Using the  supersymmetric version of the FZZ duality \refs{\fzz , \horikap, \tong}  a geometric understanding has been gained as the superstrings propagating on the cigar -- the supercoset  $SL_{2}(\IR)/U(1)$ \wittenbh\ -- tensored with flat $d$-dimensional spacetime.

Recently, heterotic strings in two dimensions with a linear dilaton profile were studied in \refs{\larsei , \SeibergNK , \DavisQI}. Previous work on heterotic strings in backgrounds with non-trivial dilaton profiles has been done in\foot{In particular, the type of models discussed in detail in this paper were anticipated in \deAlwisPR .} \refs{\deAlwisPR,  \McGuiganQP, \GiveonHM,  \GiveonZZ, \HellermanZM }. 
The purpose of this paper is to construct from the worldsheet non-critical heterotic superstrings propagating in various spacetime dimensions which interpolate -- as in the type II case -- between ten and two dimensions\foot{We shall focus on the features special to the heterotic models in the simplest class of non-critical superstrings -- they can be generalized to other theories involving {\it e.g.} minimal models, corresponding to $5$-branes wrapping other singular CY's as in \kutgiv .}. 

The theories we construct can be thought of as heterotic strings propagating on the cigar times flat spacetime. An equivalent description following \kutgiv\ is heterotic strings in 
the near-horizon limit of wrapped $NS5$-branes, or equivalently the theory describing the modes of the heterotic theory near the singularities on Calabi-Yau manifolds, in a certain double scaling limit zooming in on the singularity, simultaneously taking the string coupling to zero. 

The singular CY-manifolds which are tensored with $\IR^{d-1,1}$ in this limit are described by the non-compact manifold $\sum_{i=1}^n z_i^2 =\mu$, $n=(12-d)/2$, $z_i \in \IC$. Equivalently, we have  heterotic string theory in the near-horizon background of NS5-branes with $d$ flat spacetime directions and $6-d$ directions wrapped on $\sum_{i=3}^n z_i^2 = \mu$.
In this way, the various theories interpolate between ten dimensions all the way down to two dimensions, where they connect to well-understood theories \SeibergBX\ with asymptotically linear dilaton backgrounds\foot{The theory with $\mu=0$ is singular, and corresponds on the worldsheet to a pure linear dilaton theory with a strong coupling singularity.}. Zooming in on the singularities (or equivalently, ``wrapping'' the $NS5$-branes on non-compact curves) effectively freezes some of the transverse perturbative degrees of freedom and induces a linear  dilaton profile.

Ten-dimensional heterotic theories in the background of  $NS5$-branes have been well-studied  \refs{\gavanar, \kapustin}. In addition to a non-trivial gravitational background, the presence of the heterotic $5$-brane implies a non-trivial instanton background of the corresponding gauge theory in the transverse dimensions. As in the case of many parallel heterotic $5$--branes in ten dimensions \kapustin , the double scaling limit involves taking the size of the instanton to zero along with the string coupling, keeping a certain ratio of powers of these two quantities  fixed\foot{The singular linear dilaton case corresponds to the case where the instanton is really of zero size, even in the near horizon limit.}.

\subsec{Summary of the various theories in the singular linear dilaton background}

We shall first construct supersymmetric heterotic theories on spacetimes of the form $\IR^{d-1,1} \times \CN=2$ linear dilaton. The supersymmetric right moving fields on the worldsheet are the same as the non-critical type II superstrings, and the left moving fields consist of the leftmoving part of the  non-compact bosons, as well as a lattice of the appropriate dimension. These fields should combine to produce a modular invariant theory on the worldsheet. 
We find that there are two lines of heterotic theories, one descending from the $E_{8} \times E_{8}$ theory in ten dimensions, and the other descending from the $SO(32)$ theory in ten dimensions. 
In accord with the above discussion of zooming in on singular backgrounds, we find  that some of the gauge currents of the ten-dimensional theories also get frozen in the heterotic theories, and the gauge groups arising in the various theories are subgroups of the above two of lower rank. 


For the convenience of the reader, the results regarding the gauge groups\foot{In this summary, we shall not distinguish between the SO group and the corresponding Spin groups. The details of which conjugacy classes are chosen are presented in sections 4 and 5.} and global symmetries  of the various heterotic theories are summarized here. The details of the construction are in the bulk of the paper.

{\centerline{Table 1: {\it Symmetries and Allowed gauge groups in the linear dilaton theories}}}
\medskip
\centerline{\vbox{\offinterlineskip
\hrule
\halign{\vrule # & \strut\ \hfil #\ \hfil & \vrule # & \ \hfil $#$ \hfil \ & \vrule # & \ \hfil $#$ \hfil \ & \vrule # & \ \hfil $#$ \hfil  \ & \vrule # & \ \hfil $#$ \hfil  \ & \vrule #\cr
height3pt&\omit&&&&&&&&&\cr 
&{\bf Theory}&&  d=6 && d=4 && d=2 &&d=0& \cr
height3pt&\omit&&&&&&&&&\cr 
\noalign{\hrule}
height3pt&\omit&&&&&&&&&\cr 
& Supersymmetry && \CN =(1,0) &&  \CN=1 && \CN=(1,0)  && \CN=1  &\cr
height3pt&\omit&&&&&&&&&\cr 
\noalign{\hrule}
height3pt&\omit&&&&&&&&&\cr 
& R symmetry && O(3)_{R} &&  U(1)_{R} && (U(1) \times \IZ_2)_R && U(1)_R &\cr
height3pt&\omit&&&&&&&&&\cr 
\noalign{\hrule}
height3pt&\omit&&&&&&&&&\cr 
& KK gauge field &&  {\bar{SU(2)}} &&   {\bar{ U(1)}} &&  {\bar{U(1)}}  &&  --  &\cr
height3pt&\omit&&&&&&&&&\cr 
\noalign{\hrule}
height3pt&\omit&&&&&&&&&\cr 
& $E_{8} \times E_{8}$ line && E_{8} \times E_{8}   &&  E_{8} \times E_{7} &&  E_{8} \times E_{6} & &  E_{8} \times SO(10) &\cr
height3pt&\omit&&&&&&&&&\cr 
\noalign{\hrule}
height3pt&\omit&&&&&&&&&\cr 
& $SO(32)$ line && SO(32)&& SO(28) \times SU(2)  && SO(26) \times U(1)  & & SO(24) \times U(1)  & \cr
height3pt&\omit&&&&&&&&&\cr}
\hrule
}}

\medskip
\leftskip=35pt
\rightskip=35pt
\baselineskip=12pt
The two lines of theories correspond to wrapped NS$5$-branes of the two kinds from the ten-dimensional point of view. The theory labelled by $d$ admits a $Poincare(d-1,1)$ symmetry for $d>0$.
The details regarding the chosen conjugacy classes are provided in the bulk of the paper. The Kaluza-Klein gauge fields\foot{These fields are KK modes in the $d$ dimensional theories arising from gravitons with one leg in the compact space. In the $d=0$ theory, both the legs of the graviton would be on the compact direction, and in fact, there is one such physical discrete state. However, this mode naturally combines with the other gauge factor to give rise to a larger gauge group. One of the entries is left empty to avoid this overcounting. } which are denoted here by barred currents arise from the supersymmetric right movers in the compact directions.

\leftskip=0pt
\rightskip=0pt
\baselineskip=18pt

\bigskip

\subsec{The various theories in the smooth cigar background}

From the worldsheet point of view, the linear dilaton theories above suffer from a strong coupling singularity. In the type II case referred to earlier, this is resolved by  turning on the $\CN=2$ Liouville operator, or considering its mirror cigar coset $SL_{2}(\IR)_{k}/U(1)$ where the level of the coset is tuned to make the worldsheet theory  free of conformal anomalies. These theories no longer suffer from strong coupling singularities, and the effective tunable string coupling at the tip of the cigar is a modulus of the theory. 

In the heterotic theories, one can turn on the same $\CN=(2,2)$ deformation  \gavanar\  -- this ensures that the CFT is a well-defined string background at tree level -- and consider the background $\IR^{d-1,1} \times  SL(2)_{k}/U(1)$. From the  point of view of gauge theory in spacetime, this is equivalent to turning on a certain instanton background which breaks some of the gauge symmetry of the singular theory. We find then that allowed gauge groups are:


{\centerline{Table 2: {\it Allowed gauge groups in the cigar theories}}}
\medskip
\centerline{\vbox{\offinterlineskip
\hrule
\halign{\vrule # & \strut\ \hfil #\ \hfil & \vrule # & \ \hfil $#$ \hfil \ & \vrule # & \ \hfil $#$ \hfil \ & \vrule # & \ \hfil $#$ \hfil  \ & \vrule # \cr
height3pt&\omit&&&&&&&\cr 
&{\bf Theory}&&  d=6 && d=4 && d=2 &  \cr
height3pt&\omit&&&&&&&\cr 
\noalign{\hrule}
height3pt&\omit&&&&&&&\cr 
& $E_{8} \times E_{8}$ line && E_{8} \times E_{7}   &&  E_{8} \times E_{6} &&  E_{8} \times SO(10) &  \cr
height3pt&\omit&&&&&&&\cr 
\noalign{\hrule}
height3pt&\omit&&&&&&&\cr 
& $SO(32)$ line && SO(28)  \times SU(2) && SO(26)  \times U(1) && SO(24) \times U(1) &   \cr
height3pt&\omit&&&&&&&\cr
}
\hrule
}}
\medskip
\leftskip=35pt
\rightskip=35pt
\baselineskip=12pt

\ndt The cigar theory has the same supersymetries  and global bosonic symmetries as those presented in Table 1. The gauge groups from the left movers are smaller, and the right moving KK gauge fields are absent. The $d=0$ case will be discussed in more detail in the following.

\leftskip=0pt
\rightskip=0pt
\baselineskip=18pt


\subsec{The moduli and the spacetime interpretation}
For large values of $k$, the heterotic $5$-branes can be understood as instantons in the low energy effective theory. For our cases, since the background has curvatures of order string scale, the approximation of gravity plus gauge theory describing the closed string physics is not a priori valid. However, we can try to compute the closed string moduli in the theories, and ask if the moduli space is that of a known low energy theory. In general, the  singular string theory has many moduli which are lifted in the smooth deformed theory. 

In the case of $d=6$ however, the smooth theory on the cigar also has surviving moduli. We shall see that these moduli are precisely those of an instanton in the gauge theory in transverse space, as in  the picture of  the wrapped\foot{For d=6, the branes are not really wrapped, and are points in the transverse space .}  $5$-branes described above.
In this case, the gauge theory lives on $\IR^{4}$ and the ADHM construction provides an explicit construction of the instantons. A  finite size instanton in the gauge groups $E_{8} \times E_{8}$ and $SO(32)$  breaks the gauge groups to $E_{8} \times E_{7}$ and $SO(28) \times SU(2)$ respectively and breaks the global $SO(4)$ symmetry to an $SU(2)$. 

There are  four zero modes of the  instanton which are singlets under the gauge groups corresponding to its size and orientation in the $\IR^{4}$. There are also zero modes charged under the gauge fields which correspond to the orientation  of the instanton in the gauge group. For the two theories, the charged moduli are doublets under the global $SU(2)$ and transform as $({\bf 1, 56})$ and $({\bf 28, 2})$ of the respective gauge groups \gavanar . We shall see that the $d=6$ string theory indeed realises these moduli as exactly marginal operators on the worldsheet. 

For the other cases, the wrapped $5$-brane picture gives a gauge theory living on a curved space, and the zero mode mode analysis is more difficult. A preliminary analysis in the string theory indicates that there are no moduli charged under the surviving gauge group in the other cases. 

\subsec{Plan of the paper}

The rest of the paper is organized as follows. In section 2, we present a quick review of the type II theories, recalling the features relevant for the heterotic analysis. In section 3, we present the worldsheet theory for a general $d$, and present the problem of the heterotic construction as finding a  lattice with specific properties \narain . In section 4 and 5, we present some details of the two lines of theories in dimensions $d=6,4,2,0$,  the construction of  modular invariant torus partition functions, the ensuing spectrum, and a discussion of the gauge and global symmetries for singular and deformed theories.  Section 4 also contains a more detailed discussion of the moduli from the worldsheet point of view, with the $d=4$ and $d=6$ cases in the $E_{8} \times E_{8}$ theory treated in particular detail. An appendix collects all the relevant characters and modular transformation formulas.

\newsec{Quick Review of the non-critical type II theories}
Type II superstring theory on backgrounds of the form $\IR^{d} \times (\CN=2 \, {\rm linear \, dilaton})/{\rm cigar}$ have been studied in detail \refs{\sameer , \kutgiv, \mizo}. We shall quickly present the relevant features of the type II theories in dimensions $d=6,4,2,0$ which shall be used to construct the heterotic theories.  The singular manifolds mentioned in the introduction correspond to $K3$ and $CY3,4,5$-folds respectively.

The worldsheet fields are $\rho, \theta, X^{\mu}, \psi^{\rho}, \psi^{\theta}, \psi^{\mu}$, $(\mu=0,1..d-1)$. All the bosons and fermions are free except for $\rho$ along which the dilaton varies linearly $g_{s} = e^{-{Q \over 2} \rho}$; this induces a background charge and makes its central charge $c_{\rho}=1+3 Q^{2}$. By $\CN=2$ supersymmetry on the worldsheet, the radius\foot{We shall work in conventions where $\alpha' = 2$.} of the coordinate $\theta$ is $R=2/Q^{2}$. To get a consistent string background, we add the $(b,c,\b,\gamma)$ ghosts and tune $Q^{2} = {8-d \over 2}$ so that the total central charge vanishes.

\subsec{d=6, or singular K3 compactifications}
For $d=6$, the boson $\theta$ is at the free fermion radius and can be written as $\psi^{\pm} = e^{\pm i\theta}$. The three fermions $\psi^{3} \equiv \psi^{\theta} , \psi^{\pm}$ make up an $SU(2)_{1}$ current algebra. 
The worldsheet theory then has $X^{\mu},  \psi^{\mu}, \, \mu=0,..5, \; \rho, \psi^{\rho}, \psi^{i}, i=\pm,3$ with a total of ten free fermions. This description makes it clear that the partition function is closely related to the ten-dimensional superstring. The type II  torus partition function is\foot{Here, we only count the operators which are non-normalizable near the weak coupling end, that are the ones local on the worldsheet \SeibergEB. We discuss this further later.}:
\eqn\sixdtypeIIpfn{
{\bf Z} = V_{8} \int {  d^{2} \tau \over \tau_{2}^{2}} {1 \over \tau_{2}^{5/2}} {1 \over |\eta(\tau)|^{10}}  \times 
\left| {\vartheta_{00}^{4}(\tau) \over \eta^{4}(\tau)} - {\vartheta_{01}^{4}(\tau) \over \eta^{4}(\tau)} - {\vartheta_{10}^{4}(\tau) \over \eta^{4}(\tau)}  \right |^{2}
}

In the linear dilaton theory, there is an $SU(2)_{L} \times SU(2)_{R}$ symmetry and there are six-dimensional (massless) gauge fields in the spectrum for the corresponding currents. Only a diagonal (global) $SU(2)$ is a true symmetry. This can be understood by the presence of non-zero correlators in the theory on the coset with non-zero net charge under the broken $SU(2)$. In the dual ($\CN=2$ Liouville theory), this is understood by the fact that only the conserved symmetries commute with the interaction term in the Lagrangian. There are also 9 scalar massless fields in the ${\bf 3} \times {\bf 3} = 1 + 3 + 5$ of the diagonal, out of which only the $1+3$ are truly marginal at second order in conformal perturbation theory \sameer , and the moduli space is  $\IR^{4}/Z_{2}$. These calculations match with the geometric picture of two NS$5$-branes separated in the transverse four directions (the coulomb branch).

\subsec{d=4, or singular CY3 compactifications}
In this case, the boson lives on the self dual radius, $R=\sqrt{2}$, and there are two modular blocks made up of the   $SU(2)_{1}$ characters and the four fermions remaining in light cone gauge. We first define:
\eqn\defZkfourd{
Z^{4d}_{k}(\tau)  = \left({\vartheta_{00}^2(\tau) \over \eta^2(\tau)} - e^{\pi i k} {\vartheta_{01}^2(\tau)\over \eta^2(\tau)}\right) {\vartheta_{k\,0}(2 \tau) \over \eta(\tau)} - {\vartheta_{10}^2(\tau)\over \eta^2(\tau)}  {\vartheta_{k+1\,0}(2 \tau) \over \eta(\tau)}
}
\ndt The type II partition function is
\eqn\fourdtypeIIpfn{
{\bf Z} = V_6 \int {d^2 \tau \over \tau_2^2} {1 \over \tau_2^{3/2}} {1\over |\eta(\tau)|^6}  \sum_{k=0}^{1} \left| Z_{k}^{4d}(\tau) \right|^{2}
}
It can be checked that the characters $Z^{4d}_{k}(\tau)$ have the same modular transformation properties as those of $E_{7}$. 

The linear dilaton theory has a $U(1) \times U(1)$, which is broken to a global $U(1)$ as above. The conserved and broken $U(1)$ correspond to momentum and winding around the cigar.

\subsec{d=2, or singular CY4 compactifications}
Here, the boson lives on a circle of radius $R=2/\sqrt{3}$ and there are three modular blocks made up this boson and the two remaining fermions. We first define:
\eqn\defZktwod{
Z^{2d}_{k}(\tau) = {\vartheta_{00}(\tau) \vartheta_{{2k\over3}0}(3\tau)\over \eta^2(\tau)}  - 
e^{2 \pi i k \over 3}{\vartheta_{01}(\tau) \vartheta_{{2k\over3}1}(3\tau)\over \eta^2(\tau)} - 
{\vartheta_{10}(\tau) \vartheta_{1+{2k\over3}\,0}(3\tau)\over \eta^2(\tau)} 
}
The type II partition function is the following:
\eqn\twodtypeIIpfn{
{\bf Z} = V_4 \int {d^2 \tau \over \tau_2^2} {1 \over \tau_2^{1/2}} {1\over |\eta(\tau)|^2}  \sum_{k=0}^2 \left| Z_{k}^{2d}(\tau) \right|^{2}
}
The characters $Z^{2d}_{k}(\tau)$ have the same modular transformation properties as those of $E_{6}$. 

The $U(1)$ symmetries are the same as the $d=4$ case.

\subsec{d=0, or singular CY5 compactifications}
In this theory, there are no transverse oscillators, and the only remaining modes are the momemtum and winding of the boson which lives at the inverse of the free fermion radius $R=1$. The type II partition function is\foot{In this case, one subtlety is that the Dirac equation in spacetime has to be imposed later by hand.} 
\eqn\zerodtypeIIpfn{
{\bf Z} = V_2 \int {d^2 \tau \over \tau_2^2}  \sum_{m,w=-\infty}^\infty \exp\left({-\pi |m-w\tau|^2\over 2 \tau_2}\right) 
}

\newsec{Supersymmetric Heterotic theories}

\subsec{Construction using lattices}

We shall take the right movers to be the same as the type II non-critical superstring. To get the heterotic string, we replace the left moving fermions and superghosts by a lattice with the same central charge $c_{L} = 11+ {d +2 \over 2} = 12 + {d \over 2}$. Along with the right moving fermions, and the left and right moving compact boson $\theta$, we have then lattices of the form $\Gamma_{13+{d\over 2},{1+{d\over 2}}}$ which describe the string theory in light cone gauge.\foot{We can also use covariant lattices \lust , but since we always have two flat light cone directions for $d>0$, we can always choose this gauge. We remark on the $d=0$ case below.} To get a modular invariant one-loop amplitude, the condition becomes that $\Gamma_{13+{d\over 2},{1+{d\over 2}}}$ is an {\it odd self-dual lattice}.

To guarantee spacetime susy, we demand that the right moving partition function is exactly the same as in the type II string\foot{This has an interpretation as the singular CY/wrapped NS$5$-brane geometries, but is not necessarily the only choice. A more general problem is to classify all such lattices which have spacetime susy.}. 
Then, the problem of constructing supersymmetric theories which are modular invariant at one loop reduces to finding a lattice $\G_{13 + {d \over 2}}$ whose  characters transform in exactly the same way as the right movers of type II.\foot{Note that this problem is different from the problem of genuine compactifications to $d$ dimensions \narain\  because the left and right movers of the ``internal'' lattice are tied together by the dilaton direction.}  These lattices have a $13 + {d \over 2}$ dimensional branch of moduli space given by $SO(13 + {d \over 2},1)/SO(13 + {d \over 2})$ transformations of the above one.

From the type II analysis, we know that we need to replace the left movers with a lattice $\G_{13+{d \over 2}}$ whose characters transform\foot{This is strictly true for $d=6,4,2$ because of the subtlety for the $d=0$ case mentioned above. However, it is true for the $d=0$ case as well that the lattices follow the same pattern and they transform as the rank $5$ group $SO(10)$. Indeed as in \larsei , a covariant formulation would treat all the cases on the same footing.} as $E_{5+{d \over 2}}$.  Under the $\CS$ modular transformation $(\tau \to -{1 \over \tau})$, they behave the same way as the characters of $A_{6-d \over 2}$. Assuming that these theories have a ten-dimensional origin, where the gauge groups can only be $E_{8} \times E_{8}$ and $SO(32)$, we can reduce the problem to the following: embed $A_{6-d \over 2}$ in these two groups, and within the maximal commuting subgroups, find conjugacy classes which have the correct transformation under $T:\tau \to \tau + 1$ to match the right movers. In this way, we find the results in Table 1.

\subsec{GSO projection, Supersymmetry}
An equivalent way to ensure spacetime supersymmetry is to use a  GSO type projection on the states in the theory for both the left and right movers. For all the theories, we have made a list of the low lying vertex operators using the GSO projection, in the following two sections. To do this, one demands locality of the operators with respect to the supercharges built using the supersymmetric right movers of the string worldsheet. The fields $S=e^{-{\varphi \over
2} +i{\phi\over 2}}$ and $\bar S=e^{-{\varphi \over 2} -i{\phi\over 2}}$ are each multiplied by spin fields which are spinors of $Spin(d)$ (for $d/2$ even these are conjugate spinors and for $d/2$ odd they are the same spinor).

The algebra obeyed by the supercharges is the minimal\foot{In the type II theories, there were an equal number of supercharges from the left movers.} Poincare superalgebra in $d-1,1$ dimensions.
For $d/2$ odd or even, one has respectively 
\eqn\susyalg{
\{\CS_\alpha,\bar \CS_\beta\}= 2\gamma^\mu_{\alpha\beta}P_\mu, \qquad {\rm or} \qquad 
 \{\CS_\alpha,\bar \CS_{\dot \beta}\}= 2\gamma^\mu_{\alpha\dot\beta} P_\mu.
}
The symmetry generator $P^\theta$ corresponding to the translation around the cigar is an $R$ symmetry.
\eqn\pssbaralg{
  [P^\theta, \CS_\alpha]=\half \CS_\alpha, \qquad [P^\theta, {\bar \CS}_{\dot \alpha}]=-\half {\bar \CS}_{\dot \alpha}. 
}

\subsec{Heterotic theories on the cigar}

The linear dilaton theories have a strong coupling singularity; in the type II case, these are resolved by turning on the sine-Liouville operator, or equivalently, by considering the manifold to be an $SL(2)_{k}/U(1)$ coset. Because of $(2,2)$ supersymmetry on the worldsheet, these theories are exact solutions to string equations of motion at $g_{s}=0$. 
In the heterotic case as well, we shall focus on $(2,2)$ compactifications which means that we shall turn on the same types of operators to resolve the strong coupling singularity. This ensures that the resulting CFT's are well-defined string backgrounds.

\vskip0.5cm

\ndt  {\it  The spectrum and exact symmetries }

Not all the symmetries of the singular linear dilaton theory are those of the smooth cigar theory. The operator that is turned on in the worldsheet breaks some of the gauge and global symmetries of the singular theory, as in the type II case. To understand which symmetries remain conserved in the full theory, one needs to check whether the correlation functions in the coset are invariant under the transformation under consideration. The duality with the $\CN=2$ Liouville theory gives an easier method, which is to demand that the symmetry generators commute (have no simple pole in the OPE) with the  $\CN=2$ Liouville winding interaction  
\eqn\interactterm{
\delta S = \int d^{2} z \; (\psi_{\rho}-i \psi_{\theta})e^{{1 \over Q}(- \rho + i  \theta)} (\T \psi_{\rho}+ i \T \psi_{\theta})e^{{i \over Q}(-{\T \rho} - i \T \theta)} + c.c. 
}
In this manner, we find the gauge fields which carry the conserved symmetries in the full theory that are presented in Table 2.

It is worth pointing out that our SCFT admits a modulus -- the coefficient of the above interaction --  which can be varied without affecting the conformal dimensions of the operators. This is as in Liouville theory. The point is that one can compute the one-loop partition function at weak coupling where the potential is highly suppressed and one use the linear dilaton theory. This is in contrast with compact SCFTs, where on turning on a modulus, the spectrum changes, and solving the string tree level theory already tells us about which particles are massless and which gauge symmetries are exact. 

The comments in the above paragraph are illustrated better by an example which is more familiar and is present in the type II and bosonic theories -- that of the momentum and winding ``symmetries'' on the cigar. The physical spectrum includes massless gauge fields corresponding to both of these symmetries, but the gauge field carrying winding charge does not commute with the $\CN=2$ Liouville interaction. In the language of the dual coset theory, the correlation functions carry non-zero charge under this symmetry.

Another point worth noting is the following. Each oscillatory state of the string can be interpreted as in flat space as a mode of a field living on the $d+2$ dimensional geometry. One must keep in mind however that the circle is small in string units and it is more natural to organize the spectrum as fields living in the $d+1$ non-compact directions. 
However, the supersymmetry group of the theories is $d$ dimensional Poincare superalgebra\foot{For $d=0$, there are no transverse dimensions and one can organize the spectrum as a set of fields on the cylinder or cigar.} \susyalg , \pssbaralg , and it is more natural to classify the observables as a set of representations of the $d$ dimensional algebra labeled by a continuous momentum in the dilaton direction. This applies also for the type II theories. As an example (this is explicitly constructed in the next section), for $d=4$ at the lowest level, there is a vector particle living in five dimensions, and its supermultiplet contains also fermions and {\it one} real boson.  This multiplet structure is that of an off-shell vector multiplet in four dimensions.\foot{This might look a little strange from the point of view of four dimensional particle physics, but these modes are particles in five dimensions, not four.} 

\vskip0.5cm

\ndt {\it The moduli}

The last general comment has to do with the exact moduli of the theory. These are a little more complicated than the analysis of the symmetries, the reason having to do with the fact that the moduli or the massless scalars of the $d$ dimensional theory correspond to string modes that are localized near the tip of the cigar, in other words {\it normalizable} modes. One way to analyze this problem is to include the discrete states and write a modular invariant partition function. For the type II case, this was done recently in \EguchiIK . 

To analyze the problem using an interacting worldsheet Lagrangian, we proceed as follows. We assume that worldsheet theory is the $\CN=2$ Liouville interaction \interactterm . We know that this generates at the quantum level at least one other operator, {\it i.e.} the cigar interaction\foot{Note that this latter operator is always normalizable. The former is a non-normalizable (local) operator for $d=0,2,4$.} $\delta S = \int d^{2}z \CL^{cig} \T{\CL^{cig}}  $ where $\CL^{cig} = (\psi^{\rho} \psi^{\theta} + Q \p \theta)e^{-Q \rho}$. Since the moduli fields are normalizable, they have a dependence on $\rho$ which look like $e^{(p-Q/2)\rho}, \, p<0$. Even if we ignore the fact that these operators are non-local and try to take a formal OPE with the interaction, one generates in general an infinite number of terms which are more and more normalizable. 

We shall defer the algebraic analysis of the discrete states in these models and adopt the following strategy. Assuming a worldsheet Lagrangian consisting of the $\CN=2$ Liouville and cigar interactions, we define an operator to be a modulus in spacetime if its OPE with the above interactions is non-singular. For $d=6$, our results for the moduli match the spacetime interpretation of the near-horizon limit of $k=2$ heterotic five-branes as one instanton\foot{Although $k=2$, only the relative moduli of two instantons survive in the near horizon limit  \gavanar, \kapustin .} of the relevant gauge group living in the transverse $\IR^{4}$.  For the other values of $d$ (we shall lay out the details for $d=4$ in the next section), we find that there are no moduli -- the potentially massless scalars all have a non-zero OPE with the interactions in a way that gives mass to the scalars. 
Based on this, we would expect the relevant codimension four objects in spacetime, arising from wrapped five-branes on the non-trivial curves mentioned in the introduction have no zero modes. Because of the wrapping however, the spacetime setup as instantons on a certain transverse space is not easy to solve.

\newsec{The $E_{8} \times E_{8}$ line of theories}

In this section, we shall work out in detail the aspects of the various theories with the above gauge group. For one of the cases ($d=4$), we present a detailed analysis of the spectrum, gauge groups in the singular linear dilaton theory and smooth cigar theory, and moduli.  For the other cases, we only list the aspects that are particular to that case. For the case $d=6$ which has extended worldsheet supersymmetry, we present a discussion of the moduli in more detail -- in particular, the  matching with the spacetime picture of $E_{8} \times E_{8}$ instantons. In the next section, we shall do the same for the $SO(32)$ line of theories, highlighting the new aspects. 

The general setup is the following. The worldsheet fields contain as in the type II case, the left and right moving bosons $\rho, \theta, X^{\mu}$ and the $(b,c)$ ghosts. On the supersymmetric rightmoving side, there are the fermions $\psi^{\rho}, \psi^{\theta}, \psi^{\mu}$, $(\mu=0,1...d-1)$ and $(\b,\gamma)$ ghosts. On the leftmoving side, we have the fermions $\T \lambda^{i},\,(i=1,2... 22+d)$ as well as $\T\psi^{\rho}, \T\psi^{\theta}$. Taking into account the slope of the dilaton $\rho$, the total central charge on both sides is then zero. 

In all the theories in this section, the leftmoving fermions are split into two groups -- $\T \lambda^{i}, i = 1..16$ which form one of the $E_{8}$ factors, and do not interact with the rest, and the rest of the $6+d$ fermions which combine with $(\psi^{\rho}, \psi^{\theta})$, and the compact boson $\theta$ to give the other factor.

\subsec{d=4}
For the linear dilaton theory, based on the rightmoving part of the type II theory, we look for a lattice  with rank 15 whose characters transform exactly like $E_{7}$. 
The $E_{8}$ lattice has only one conjugacy class $(0)$, and so the $E_{8} \times E_{7}$ lattice has the required properties. It has two conjugacy classes transforming like $E_{7}$. 
The modular invariant partition function\foot{The characters used here and below are all characters of the corresponding current algebra at level one. A summary of these is given in the appendix.} is thus given by:
\eqn\fourdhetpfn{
{\bf Z} = V_6 \int {d^2 \tau \over \tau_2^2} {1 \over \tau_2^{3/2}} {1\over |\eta(\tau)|^6}  \sum_{k=0}^{1} \left(Z_{k}^{4d}(\tau)\right)^{*}  Z_{k}^{E7}(\tau)
}

To construct the vertex operators, we first construct the building blocks for the gauge currrents:
\eqn\ggecurfourd{\eqalign{
E_{8}  \; {\rm gauge \; current} \; A^{\a \b}: & \quad  \T \lambda^{i}  \T \lambda^{j}, \; (i,j = 1,2..16); \quad \T \sigma \in {\bf 128} \; {\rm  of}\;  SO(16) . \cr
E_{7}  \; {\rm gauge \; current}  \; A^{ab}: & \quad  \T \lambda^{i}  \T \lambda^{j},  \quad \T \lambda^{i}  \T \psi^{\rho,\theta} , \quad  \T \psi^{\rho} \T \psi^{\theta} , \quad (i,j = 17,..26); \cr
 &  \quad \T j^{\pm} \equiv e^{\pm i{\rt2}{\T \theta}}, \quad \T j^{3} \equiv \pb \T \theta.\cr
& \quad   \T \Sigma e^{\pm {i \over 2} \left( \t H + \rt2 \t \theta \right)} , \quad \T \Sigma \in {\bf 16} \; {\rm  of}\;  SO(10) \cr
& \quad  \bar{ \T \Sigma} e^{\pm{i \over 2} \left( \t H - \rt2 \t \theta\right)}, \quad \bar{\T \Sigma} \in {\bf \bar{16}} \; {\rm  of}\;  SO(10) . \cr
}}
The $SO(16)$ and  $SO(10)$ above rotate the fermions $\left(\T \lambda^{i},\, i=1..16 \right)$ and   $\left(\T \lambda^{i},\, i=17..26 \right)$ respectively;  $\T \sigma$ and $(\T \Sigma, \bar{\T \Sigma})$ are the spin fields of the corresponding set of fermions.

We can now put together the left and right movers, imposing the BRST condition coming from the $\CN=1$ gauging of the right movers, along with the GSO condition implied in the partition function  \fourdhetpfn . We can then classify the operators as spacetime fields on the the non-compact part of the space $\IR^{4} \times \rho$. Since there is no translational invariance along $\rho$, we classify them by the remaining symmetry which is that of the $\CN=1$ Poincare superalgebra in $d=4$ with the R symmetry of translations around the circle; as well as the gauge symmetries above. In general, there will be a label $p$ on these operators corresponding to the profile in the dilaton direction, below we shall restrict to the massless modes in $\IR^{4}$ which  forces $p^{2} = \half$.  We shall also restrict to the operators obeying the Seiberg bound, i.e. the non-normalizable branch.  

\ndt The asymptotic vertex operators for the  massless ($k^{\mu} k_{\mu} = 0$) states are:
\eqn\masslessfourd{\eqalign{
{\rm Graviton:}  & \quad  e^{-\varphi} \psi^{\mu} \pb X^{J} e^{ik_{\mu} X^{\mu}}  \cr
E_{8}  \; {\rm gauge \; field:}  & \quad  e^{-\varphi} \psi^{\mu} A^{\a \b}  e^{ik_{\mu} X^{\mu}}\cr
E_{7}  \; {\rm gauge \; field:}   & \quad  e^{-\varphi} \psi^{\mu} A^{ab}  e^{ik_{\mu} X^{\mu}}\cr
{\rm E_{8} \times E_{7} \; Adjoint \; scalar:} & \quad  e^{-\varphi} \psi^{\theta} \times (as \; above) \cr
&  \mu = 0,1..3.
}}
Note that the states with $\psi^{\rho}$ excitations in the massless sector have been thrown out by the BRST constraint. 

By acting on the above operators with the supercharges, we can figure out the susy multiplets. The susy algebra is $\CN=1$ Poincare susy in $d=4$, with the $U(1)$ R symmetry given by the momentum around the cigar. Asymptotically, the supercharges do not depend on the coordinate $\rho$. We find that the vector and the one scalar in \masslessfourd\ are in a off-shell vector multiplet of the above algebra. Combined as a vector multiplet in $d=4$, the scalar is in the position of the D-term field for the above vector. 

\vskip0.5 cm

\ndt {\it Comment on extended symmetry algebra:} In the type II theory in $d=4$, in spite of the fact that the boson lives on a circle of self-dual radius, there was no symmetry enhancement of the $U(1)$ of the circle to $SU(2)$, since the $SU(2)$ currents were not physical in the super-linear dilaton theory. In the heterotic case, the left-moving $SU(2)$ currents from the boson are indeed physical and they are required to complete the gauge group to $E_{7}$. In the $SO(32)$ line, this $SU(2)$ is actually seen as a direct product with the rest of the $SO(28)$ currents, as we shall see in the next section. 

\vskip0.5 cm

\ndt {\it Interacting theory}

All the above operators \masslessfourd\ are present in the partition function \fourdhetpfn\ of the linear dilaton theory. Since the partition function of the cigar theory which is independent of $g_{s}$ can be written in the weak coupling region, the above operators are also present in the cigar theory. In the coset theory, there are also other modes in the discrete representations which are not present in the partition function \fourdhetpfn . 

The massless operators above remain massless at $g_{s}=0$ in the interacting theory. However, as in the type II case, not all the symmetries of the linear dilaton theory carry over to the cigar, {\it i.e.} in the presence of  the  $\CN=2$ Liouville (winding) interaction:
\eqn\interacttermfourd{
\delta S = \int d^{2} z (\psi_{\rho}-i \psi_{\theta})e^{{1 \over \rt2}(- \rho + i  \theta)} (\T \psi_{\rho}+ i \T \psi_{\theta})e^{{i \over \rt2}(-{\T \rho} - i \T \theta)} + c.c. 
}
The gauge fields which commute with the interaction and thus correspond to conserved symmetries in the coset theory are (all operators have $k^{\mu} k_{\mu} = 0$):
\eqn\conservedfourd{\eqalign{
E_{8}  \; {\rm gauge \; field:}  & \quad  e^{-\varphi} \psi^{\mu} A^{\a\b}  e^{ik_{\mu} X^{\mu}}.  \cr
E_{6}  \; {\rm gauge \; field:}   & \quad  e^{-\varphi} \psi^{\mu} \t \lambda^{i}  \t \lambda^{j}  e^{ik_{\mu} X^{\mu}} , \; (i,j = 17,..26); \cr
& \quad  e^{-\varphi} \psi^{\mu} \T \Sigma e^{\pm {i \over 2} \left( \t H +  \rt2 \t \theta \right)}  e^{ik_{\mu} X^{\mu}}, \quad \T \Sigma \in {\bf 16} \; {\rm  of}\;  SO(10); \cr
& \quad  e^{-\varphi} \psi^{\mu} (\T \psi^{\rho} \T \psi^{\theta} - \rt2 \,  \pb \T \theta) e^{ik_{\mu} X^{\mu}}. \cr
}}

Apart from these gauge fields, we could also have {\it global} symmetries which commute with the interaction. These have both left and right moving components on the worldsheet. In this case, we only have the momentum $U(1)$ symmetry generated by $(e^{-\varphi} \psi^{\theta}, \pb \T \theta)$.

\vskip 0.5 cm

\ndt {\it Moduli}

For the moduli, from the supersymmetric right movers, the only scalar operator which commutes with the interaction are the right moving part of the interaction terms themselves  $\CL^{\pm} = e^{\pm i (H + {i \over \rt2 }\theta)-{1 \over \rt2} \rho} $  where $e^{\pm iH} = (\psi_{\rho} \pm i \psi_{\theta})$, and $\CL^{cig} = (\psi^{\rho} \psi^{\theta} + \rt2 \p \theta)e^{-\rt2 \rho}$. This fixes the momentum in the $\rho$ direction, and we can then look for left moving operators to combine with.  

Firstly, of course we have the interaction terms themselves which are singlets under all the gauge groups, the only normalizable one is the cigar interaction
\eqn\modulifourdlast{
{\bf 1}\; {\rm of} \; {\bf E_{6}:}  \quad  \CL^{cig} \T \CL^{cig} 
}
We find that all the other operators have a non-trivial OPE with the interaction terms. For example, with the rightmover  $\CL^{cig}$, one can combine leftmovers to get the following list of scalars in the (anti)fundamental of $E_{6}$. We write them below in the representations of $SO(10)$ generated by the rotations of the fermions $\left( \lambda^{a}, a= 17... 26 \right)$:
\eqn\modulifourdmore{\eqalign{
   &  \quad  \CL^{cig} \; \T \lambda^{a} (\T\psi_{\rho} + i \T\psi_{\theta}) e^{-{\rt2} \T \rho}, \quad {\bf 10} \; {\rm  of}\;  SO(10); \cr
{\bf 27} \; {\rm of} \; {\bf E_{6} :} \quad \quad  &  \quad  \CL^{cig} \; \T  \Sigma e^{- {i \over 2}(\T H + {\rt2 }\T \theta)- {\rt2} \T \rho}, \quad {\bf 16} \; {\rm  of}\;  SO(10);   \cr
   &  \quad  \CL^{cig} \; e^{ i{\rt2}{\T \theta}- {\rt2} \T \rho}, \quad {\bf 1} \; {\rm  of}\;  SO(10).   \cr
}}
and their complex conjugates which give ${\bf {\bar{27}}}$. 
The residue in the pole of their OPE's with the cigar interaction term is proportional to themselves, implying a Yukawa type coupling; in other words, for non-zero value of the cigar interaction, the scalars are massive.

The spacetime interpretation of the background according to \kutgiv\ is a codimension four object in the $E_{8} \times E_{8}$ gauge theory on the space transverse to $z_{1}^{2}+z_{2}^{2}=0$ embedded in $\IC^{3}$.  The analysis above implies that there are no zero modes of this object.

\subsec{d=6}
Since the boson $\theta$ is at the free fermion radius $R=2$, we can fermionize it as $e^{\pm i \theta} = {1 \over \rt2} (\psi_{1} \pm i \psi_{2}) \equiv \psi^{\pm}$.  The fermion content of both the left and right movers is thus the same as the ten dimensional heterotic string and we get an $E_{8}\times E_{8}$ gauge field. There is also an $\bar{SU(2)}$ current algebra from the right movers.
The partition function\foot{Note that the partition function is not exactly the same as the ten dimensional $E_{8} \times E_{8}$ theory, {\it e.g.} the number of $\eta$ functions are different.} is:
\eqn\sixdhetpfn{
{\bf Z} = V_{8} \int {  d^{2} \tau \over \tau_{2}^{2}} {1 \over \tau_{2}^{5/2}} {1 \over |\eta(\tau)|^{10}}  \times 
\left( {\vartheta_{00}^{4}(\tau) \over \eta^{4}(\tau)} - {\vartheta_{01}^{4}(\tau) \over \eta^{4}(\tau)} - {\vartheta_{10}^{4}(\tau) \over \eta^{4}(\tau)} \right)^{*} \left(Z^{E8}(\tau)\right).
}

The interaction term in this case is a little special, as seen in the type II theory \sameer . The free theory has a $SU(2)_{2}$ current algebra which rotates the three fermions $\psi_{i}$ on the left and the right. The group $SO(4) = SU(2)_{L} \times SU(2)_{R}$ is a symmetry of the two coincident NS5-branes. The rightmoving and leftmoving cigar  and $\CN=2$ Liouville interactions fall into a triplet of the respective $SU(2)_{2}$ current algebra. The full interaction is then a singlet under a chosen $SU(2)$ subgroup of the $SO(4)$ and transforms under the three broken generators. The choice of this $SU(2)$ (which represents the remaining symmetry of two parallel separated five-branes) plus the overall scale of separation represent the four exact moduli.

Now, let's say that the conserved $SU(2)_{cons}$ is the diagonal one\foot{It is not difficult to write down the interaction for a general conserved $SU(2)$ subgroup of the $SO(4)$ \sameer .}. After the above fermionization, the $\CN=2$ Livouille and cigar interactions can be expressed in an $SU(2)_{2}$ covariant manner in terms of  the fermion bilinears $j_i e^{-\rho} \equiv (\psi_\rho\psi_i - \half \epsilon_{ijk}\psi_j\psi_k)e^{-\rho}$. 
The interaction is then given by:
\eqn\sixdinterac{
\delta S = \int d^{2} z j_{i} \T{j_{i}}  e^{-\rho - \T \rho}. 
}

In the heterotic theory, after turning on the above interaction, the $E_{8} \times E_{8}$ gauge currents split naturally into $E_{8} \times E_{7}$ which commute with the interaction, an $SU(2)_{L}$ under which the interaction terms fall into a triplet, and the rest which have fundamental charge under at least one of the two. This $E_{8} \times E_{7} \times SU(2)_{L}$ is a maximally commuting subgroup of $E_{8} \times E_{8}$.

The $SU(2)_{L}$ currents mentioned above under which the interaction terms are a triplet are simply the $\T j_{i}$ of above\foot{Note that these currents are not physical in the type II theory, in which  the  $SU(2)_{L}$ is generated by $\half \epsilon_{ijk} \T \psi_{j} \T \psi_{k}$.}.  The gauge fields which commute with the interaction are built out of the following currents (All the corresponding operators have $k^{\mu} k_{\mu} = 0$). 
\eqn\conservedsixd{\eqalign{
E_{8}  \; {\rm gauge \; currents} \, A^{\a \b}: & \quad   \T \lambda^{i}  \T \lambda^{j}   , \,  i,j = 1,2..16; \quad \T \sigma \in {\bf 128} \; {\rm  of}\;  SO(16) .  \cr
E_{7}  \; {\rm gauge \; currents}  \, A^{ab}: & \quad   \T \lambda^{i}  \T \lambda^{j},  \,  i,j = 17...28, \cr
& \quad    \T \Sigma  \T \tau ; \quad \T \Sigma \in {\bf 32} \; {\rm  of}\;  SO(12) ,\quad \T \tau \in {\bf 2} \; {\rm  of}\;  SO(4).  \cr
& \quad  (\T \psi^{\rho} \T \psi^{i} +  \,  \half \epsilon^{ijk} \T \psi_{j} \T \psi_{k}) , \,  i=1,2,3. \cr
}}
The $SO(16)$, $SO(12)$ and the $SO(4)$ above rotate the fermions $\left( \T \lambda^{i},\, i=1..16 \right)$, $\left(\T \lambda^{i},\, i=17..28\right)$ and $\left(\t \psi^{\rho}, \psi^{i}\right)$ respectively;  $\T \sigma $, $\T \Sigma$ and  $\T \tau$ are the spin fields of the corresponding set of fermions.

\vskip0.5cm

\ndt {\it The Moduli}

The currents of $E_{8} \times E_{8}$ which do not commute with the interaction are given below.  Under the $E_{8} \times E_{7}\times SU(2)_{L}$, these fall into a ${\bf (1,56, 2)}$ which we call $M^{\a}$ (suppressing the $E_{8} \times E_{7}$ index), and a ${\bf (1,1,3)}$. 
\eqn\sixdmodulistep{\eqalign{
 {\bf (1,56, 2):} \quad  & \T{\bar \Sigma}  \, \T{\bar \tau},  \quad \T \Sigma \in {\bar{\bf 32}} \; {\rm  of}\;  SO(12) ,\quad \T \tau \in {\bar {\bf 2}} \; {\rm  of}\;  SO(4);  \cr
  & \T \lambda^{a} [(\psi^{\rho} \pm i\psi^{\theta}) + (\psi^{1} \pm i \psi^{2})],  \quad a=17.. 28;   \cr
  & \T \lambda^{a} [(\psi^{\rho} \pm i\psi^{\theta}) - (\psi^{1} \pm i \psi^{2}) ],  \quad a=17.. 28.  \cr
 {\bf (1,1,3):} \quad & \T j_{i}, \quad  i=1,2,3. \cr
}}

Now, the last step to find out which are the true moduli is to tensor the left and right movers together to produce scalars in spacetime and ask which operators commute with the interaction. The conserved symmetries of the theory are $E_{8} \times E_{7}$ gauge symmetry and the diagonal part of the product of the above $SU(2)_{L}$ and the $SU(2)_{R}$ coming from the supersymmetric side. 

The fields which are moduli in the free theory are got by combining the currents $M^{\a} e^{-\T \rho}$ and $\T j^{i} e^{-\T \rho}$ using the above leftmovers, and the three worldsheet $\CN=2$ invariant rightmovers $j^{i}e^{-\rho}$.

The problem of which are the true moduli in the interacting theory that are neutral under the gauge group has been solved already because the answer is the same as in the type II case  \sameer . 
The interaction $j^{i} \T j^{j} e^{-\rho - \T \rho}$ are in a ${\bf 3 \times {\bar 3}}$ of the $SU(2)_{L} \times SU(2)_{R}$ and under the diagonal conserved $SU(2)_{cons}$, they fall into a ${\bf 1+3+5}$. In second order of  conformal perturbation theory,  only the ${\bf 1 + 3}$ are true moduli. 

The fields which are charged as the fundamental under $E_{7}$, transform under $SU(2)_{L} \times SU(2)_{R}$ as ${\bf 2 \times 3}$ and under the diagonal conserved $SU(2)_{cons}$, they fall into ${\bf 2 + 4}$.  Among these six fields, it can be checked that the OPE of the ${\bf 4}$ with the interaction \sixdinterac\ gives  a Yukawa coupling effectively making those fields massive  whereas  the ${\bf 2}$ commute with the interaction and remain marginal.  

The final result  is that the true symmetries of the theory is $E_{8} \times E_{7}$ gauge and $SU(2)_{cons}$ global. The true moduli  transform under these groups as ${\bf (1,1,1)+}$ ${\bf (1,1,3) +} $ ${\bf (1,56,2)}$. As mentioned in the introduction, these fields have the interpretation as being the zero modes of a finite size instanton in the near horizon limit of $k=2$ heterotic $5$-branes.

\subsec{d=2}
The characters of $E_{8} \times E_{6}$ transform like those of $E_{6}$ and has the required rank 14. The partition function is given by:
\eqn\twodhetpfn{
{\bf Z} = V_4 \int {d^2 \tau \over \tau_2^2} {1 \over \tau_2^{1/2}} {1\over |\eta(\tau)|^2}  \sum_{k=0}^2  \left( Z_{k}^{2d}(\tau) \right)^{*}   Z^{E6}_{k}(\tau).
}
The gauge fields which commute with the interaction and thus correspond to conserved symmetries in the coset theory are (all operators have $k^{\mu} k_{\mu} = 0$):
\eqn\conservedtwod{\eqalign{
E_{8}  \; {\rm gauge \; currents:} \, A^{\a \b}: & \quad   \T \lambda^{i}  \T \lambda^{j} , \,  i,j = 1,2..16; \quad \T \sigma \in {\bf 128} \; {\rm  of}\;  SO(16) .  \cr
SO(10)  \; {\rm gauge \; currents:} \, A^{a b}:  & \quad   \t \lambda^{i}  \T \lambda^{j}, \,  i,j = 17,..24; \cr
&\quad  \T \Sigma e^{\pm{i \over 2} \left( \t H +  \sqrt{3} \t \theta \right)},\quad \T \Sigma  \in {\bf 8} \; {\rm  of}\;  SO(8);  \cr
& \quad  (\T \psi^{\rho} \T \psi^{\theta} - \sqrt{3} \,  \pb \T \theta). \cr
}}
The $SO(16)$ and $SO(8)$ above rotate the fermions $\left( \T \lambda^{i},\, i=1..16 \right)$ and $\left(\T \lambda^{i},\, i=17..24 \right)$ respectively;  $\T \sigma $ and $\T \Sigma$  are the spin fields of the corresponding set of fermions. There is a $U(1)$ global symmetry of translations around the cigar.

\subsec{d=0}
The linear dilaton theory is one of the theories explored in \larsei . This linear dilaton theory has a gauge group $E_{8} \times SO(10)$ where the gauge fields live in two dimensions. There is also a $U(1)$ global symmetry of translations around the cigar.
This theory is  special because it does not have any directions transverse to the cigar. The currents which make up the gauge field  in the linear dilaton theory are:
\eqn\ggecurzerod{\eqalign{
E_{8}  \; {\rm gauge \; current} \; A^{\a \b}: & \quad  \t \lambda^{i}  \t \lambda^{j}, \; (i,j = 1,2..16); \quad \T \sigma \in {\bf 128} \; {\rm  of}\;  SO(16) . \cr
SO(10)  \; {\rm gauge \; current}  \; A^{ab}: & \quad  \t \lambda^{i}  \t \lambda^{j},  \quad \t \lambda^{i}  \t \psi^{\rho,\theta} , \quad  \t \psi^{\rho} \t \psi^{\theta} , \quad (i,j = 17,..22); \cr
& \quad   \T \Sigma e^{\pm {i \over 2} \left( \t H +  \t \theta \right)} , \quad \T \Sigma \in {\bf 4} \; {\rm  of}\;  SO(6), \cr
& \quad  \bar{ \T \Sigma} e^{\pm{i \over 2} \left( \t H -  \t \theta\right)}, \quad \bar{\T \Sigma} \in {\bf \bar{4}} \; {\rm  of}\;  SO(6); \cr
& \quad \bar{\p} {\T \theta} . \cr
}}
The $SO(6)$ referred to above is the one rotating the fermions $\lambda_{i}, \, (i=17..22)$.

The above gauge currents can be tensored with rightmoving excitations of the type $e^{-\varphi} \psi_{\theta}$ to give a gauge field on the cylinder.  There are also other dimension $(1,1)$ operators  of the form $\lambda^{i}e^{\pm i \theta} $, but these are  the lowest elements  of a tower of states which are interpreted as scalars propagating on the cylinder \larsei . Note also that the mode $e^{-\varphi} \psi_{\theta} \bar{\p} {\T \theta} $ which is a graviton on the cylinder,  was interpreted as the Kaluza-Klein gauge field on flat space in the higher dimensional theories. 

In the interacting theory, none of the gauge field modes commute with the interaction because of the rightmoving excitation inside the cylinder like $e^{-\varphi} \psi_{\theta}$. In this case, an algebraic approach to the problem of finding the correct gauge symmetries using the exact $SL(2,\IR)/U(1)$ SCFT is probably necessary, as mentioned in section 3.

\newsec{The $SO(32)$ line of theories}

In this section, we shall repeat the analysis of the previous section for the line of theories descending from the gauge group $SO(32)$. We will be much more brief, since the analysis closely follows the previous section. 

The general setup is as before. The leftmoving lattice consists of the fermions $\lambda^{i}$, $i=1,2... 22+d$ along with $(\psi^{\rho}, \psi^{\theta})$, and the compact boson $\theta$. In the theories to follow, the GSO projection is such that all the fermions $\lambda^{i}$ transform into each other as well as $(\psi^{\rho}, \psi^{\theta}, \theta)$.

\subsec{d=4}
We need to find a lattice of rank $15$ with two characters which transform in the same way as those of $E_{7}$. Based on the previous experience, we embed $SU(2)$ in $SO(32)$, and the maximal commuting subgroup is $SO(28) \times SU(2)$. It can be checked that\foot{The notations of the various characters of $SU(2)$ are standard and are collected in the appendix.}
\eqn\fourdsohet{\eqalign{
Z_{0}^{SO(28)\times SU(2)} & = \chi_{0} (\tau) \chi^{A1}_{0} (\tau) + \chi_{S}(\tau) \chi^{A1}_{1}(\tau), \cr
Z_{1}^{SO(28)\times SU(2)} & =  \chi_{V}(\tau) \chi^{A1}_{1} (\tau) + \chi_{C}(\tau) \chi^{A1}_{0}(\tau) . \cr
}}
transform like the two characters of $E_{7}$. 
The full partition function is then:
\eqn\fdhetSOpfn{
{\bf Z} = V_6 \int {d^2 \tau \over \tau_2^2} {1 \over \tau_2^{3/2}} {1\over |\eta(\tau)|^6}  \sum_{k=0}^{1} \left(Z_{k}^{4d}(\tau)\right)^{*}  Z_{k}^{SO(28)\times SU(2)}(\tau)
}
In the low energy spectrum, the graviton will be of course present, and the gauge fields are:
\eqn\ggecurSOfourd{\eqalign{
SO(28)  \; {\rm gauge \; current}  \; A^{ab}: & \quad  \t \lambda^{i}  \t \lambda^{j},  \quad \t \lambda^{i}  \t \psi^{\rho,\theta} , \quad  \t \psi^{\rho} \t \psi^{\theta} , \quad (i,j = 1,..26); \cr
SU(2)  \; {\rm gauge \; current} \; A^{\a \b}: &  \quad \t j^{\pm} \equiv e^{\pm i{\rt2}{\t \theta}}, \quad \t j^{3} \equiv \pb \T \theta.\cr
}}
Note here that the modes involving the spinor conjugacy classes have at least dimension two, and so will not be present in the massless spectrum. 

In the interacting theory, the gauge fields that commute with the interaction are the $SO(26) \times U(1)$ gauge fields:
\eqn\ggecurSOint{\eqalign{
SO(26)  \; {\rm gauge \; current}  \; & A^{ab}:  \quad  \t \lambda^{a}  \t \lambda^{b},   \quad (a,b = 1,..26); \cr
U(1)  \; {\rm gauge \; current}  \; & A:  \quad  (\T \psi^{\rho} \T \psi^{\theta} - \rt2 \,  \pb \T \theta)  .
}}

\subsec{d=6}
In this case, we can fermionize $\theta$ as before to have $32$ free fermions $\lambda^{i}\, (i=1..28)$, and $\psi^{\rho}, \psi^{i}\, (i=1..3)$. In the linear dilaton theory, we have the ten-dimensional gauge group $SO(32)$ arising from the rotation of the $32$ free fermions, and an $\bar{SU(2)}$ current algebra from the right movers. 
The partition function is:
\eqn\sixdhetpfn{
{\bf Z_{T^{2}}} = V_{8} \int {  d^{2} \tau \over \tau_{2}^{2}} {1 \over \tau_{2}^{5/2}} {1 \over |\eta(\tau)|^{10}}  \times 
\left( {\vartheta_{00}^{4}(\tau) \over \eta^{4}(\tau)} - {\vartheta_{01}^{4}(\tau) \over \eta^{4}(\tau)} - {\vartheta_{10}^{4}(\tau) \over \eta^{4}(\tau)} \right)^{*} \left(Z^{SO(32)}(\tau)\right).
}
where
\eqn\sixdsohet{
Z^{SO(32)}(\tau) = \chi^{SO(32)}_{0} + \chi^{SO(32)}_{S} .
}

In the interacting theory, there is a global $SU(2)$ symmetry and 
the gauge currents which are conserved are:
\eqn\ggecurSOsixd{\eqalign{
SO(28)  \; {\rm gauge \; current}  \; & A^{ab}:  \quad  \t \lambda^{a}  \t \lambda^{b},   \quad (a,b = 1,..28) ; \cr
SU(2)  \; {\rm gauge \; current}  \; & A^{\a \b}:  \quad   (\T \psi^{\rho} \T \psi^{i} +  \,  \half \epsilon^{ijk} \T \psi_{j} \T \psi_{k}),   \quad (i=1..3).
}}

The analysis of the moduli is also similar to the $E_{8} \times E_{8}$ case and the exactly marginal operators  in the interacting theory fall into a ${\bf (1,1,1) + (1,1,3) + (28,2,2)}$ under $(SO(28) \times SU(2))_{gauge} \times SU(2)_{global}$.

\subsec{d=2}
We need a lattice with three characters which transform in the same way as those of $E_{6}$. It can be checked that the following three classes of $SO(26) \times U(1)$
\eqn\twodsohet{
Z^{SO(26)\times U(1)}_{k}(\tau) = {\vartheta_{00}^{13}(\tau)\over \eta^{13}(\tau)} {\vartheta_{{2k\over3}0}(3\tau)\over \eta(\tau)}  + 
e^{2 \pi i k \over 3}{\vartheta_{01}^{13}(\tau) \over \eta^{13}(\tau)} {\vartheta_{{2k\over3}1}(3\tau)\over \eta(\tau)} +
{\vartheta^{13}_{10}(\tau) \over \eta^{13}(\tau)} {\vartheta_{1+{2k\over3}\,0}(3\tau)\over \eta(\tau)} 
}
transform like the three characters of $E_{6}$. 
The full partition function is then:
\eqn\tdhetSOpfn{
{\bf Z} = V_4 \int {d^2 \tau \over \tau_2^2} {1 \over \tau_2^{1/2}} {1\over |\eta(\tau)|^4}  \sum_{k=0}^{2} \left(Z_{k}^{2d}(\tau)\right)^{*}  Z_{k}^{SO(26)\times U(1)}(\tau)
}
In the interacting theory, there is a global $U(1)$ symmetry and 
the gauge currents which are conserved are:
\eqn\ggecurSOtwod{\eqalign{
SO(24)  \; {\rm gauge \; current}  \; & A^{ab}:  \quad  \t \lambda^{a}  \t \lambda^{b},   \quad (a,b = 1,..24); \cr
U(1)  \; {\rm gauge \; current}  \; & A:  \quad  (\T \psi^{\rho} \T \psi^{\theta} - \sqrt{3} \,  \pb \T \theta)  .
}}

\subsec{d=0}
The linear dilaton theory is also one of the theories explored in \larsei . This linear dilaton theory has a gauge group $SO(24) \times U(1)$ with gauge fields living in two dimensions. The $SO(24)$ part arises from that many free fermions, and the $U(1)$ part arises from the discrete state $e^{-\varphi} \psi_{\theta} \bar{\p} {\T \theta}$. As before, we have classified this as a gauge field instead of a graviton. However, in this case, in contrast to the $E_{8}$ line, there are no other massless modes to enhance the symmetry and the gauge group remains a product of the two. 

There is a $U(1)$ global symmetry of translations around the cigar as in the other cases. For the same reason as in the $E_{8}$ line,  the analysis of the interacting theory would be better done using the exact coset algebra.

\vskip 1cm

\centerline{\bf Acknowledgements}
I would especially  like to thank E.~Gava and K.~S.~Narain for many helpful and enjoyable discussions on the above topic.  I would also like to thank B.~Acharya for discussions and useful comments on a preliminary draft, and D.~L\"{u}st for a useful discussion on covariant lattices. 

\vskip 1cm


\appendix{A}{Building Block Characters and their Modular transformations}
\ndt The definition of the $\vartheta$ and $\eta$ functions are:
\eqn\defeta{\eqalign{
 \eta(\tau)& =q^{1\over 24} \prod_{n=1}^\infty (1-q^n) \cr
 \vartheta_{ab}(\tau)&=q^{a\over 8} e^{\pi i ab/2} \prod_{n=1}^\infty (1-q^n) (1+e^{\pi i b}q^{n-(1-a)/2})(1+e^{-\pi i b}q^{n-(1+a)/2}) \cr
& = \sum_{n\in {\bf Z}} \exp \left[ 2 \pi i \left(\half(n+{a \over 2})^{2} \tau + (n + {a \over 2}){b \over 2} \right)\right]
}}

\ndt  We collect below the characters and their modular transformation formulas of the various current algebras used in the main text. They are all characters of current algebras at level one. 

\subsec{$SU(2)_{1}$ characters}
\ndt The characters used in the text  in the $d=4$  theories are those of a boson at the $SU(2)$ radius. There are two of them:
\eqn\sutwochar{\eqalign{
{\rm Scalar:} \quad \chi^{A1}_{0}(\tau) & ={\vartheta_{00}(2 \tau) \over \eta(\tau)}, \cr 
{\rm Fundamental:} \quad \chi^{A1}_{1}(\tau) &={\vartheta_{10}(2 \tau) \over \eta(\tau)}. \cr
}}
Their modular transformation properties are:
\eqn\sutwomod{\eqalign{
\chi_{k}(\tau+1) & = e^{i({k\over 2} -{1 \over 12} )\pi} \chi_{k}(\tau) \cr
\chi_{k}(-{1 \over \tau}) & = \sum_{l} {1 \over \rt2} e^{i \pi k l } \chi_{l}(\tau) \cr
}}

\subsec{$SO(2n)_{1}$ characters}
\ndt The characters of the four conjugacy classes of the $SO(2n)$ algebra are:
\eqn\fdtypetwonew{\eqalign{
\chi_{0}(\tau) & = \half {1 \over \eta^{n}(\tau)} \left( \vartheta_{00}^{n} (\tau) + \vartheta_{01}^{n} (\tau)\right) \cr
\chi_{V}(\tau) & = \half {1 \over \eta^{n}(\tau)} \left( \vartheta_{00}^{n} (\tau)- \vartheta_{01}^{n} (\tau)\right) \cr
\chi_{S}(\tau) & = \half {1 \over \eta^{n}(\tau)} \left( \vartheta_{10}^{n} (\tau)+ i^{n} \vartheta_{11}^{n} (\tau)\right) \cr
\chi_{C}(\tau) & = \half {1 \over \eta^{n}(\tau)} \left( \vartheta_{10}^{n} (\tau)- i^{n} \vartheta_{11}^{n} (\tau)\right) \cr
}}
The $T$ transformation is 
\eqn\sonT{\eqalign{
\chi_{0}(\tau+1) &= e^{-{i n \pi \over 12}} \chi_{0}(\tau) \cr
\chi_{V}(\tau+1) &= - e^{-{i n \pi \over 12}} \chi_{V}(\tau) \cr
\chi_{S}(\tau+1) &= e^{{i n \pi \over 6}}  \chi_{S}(\tau) \cr
\chi_{C}(\tau+1) &= e^{{i n \pi \over 6}} \chi_{C}(\tau) \cr
}}
The $S$ transformation matrx is:
\eqn\sonS{\eqalign{
\pmatrix{\chi_{0} \cr \chi_{V} \cr \chi_{S} \cr \chi_{T} \cr } (-{1 \over \tau})  =  
\pmatrix{
1& 1 &1 & 1 \cr
1& 1 &-1 & -1 \cr
1& -1 & e^{{i n \pi \over 2}}& -e^{{i n \pi \over 2}} \cr
1& -1 &-e^{{i n \pi \over 2}} & e^{{i n \pi \over 2}} \cr
} 
\pmatrix{\chi_{0} \cr \chi_{V} \cr \chi_{S} \cr \chi_{T} \cr } (\tau) 
}}

\subsec{$E_{n}$ characters}

\ndt The $E_{8}$ character is:
\eqn\charEeight{
Z^{E8}(\tau) = {\vartheta_{00}^{8}(\tau) \over \eta^{8}(\tau)} + {\vartheta_{01}^{8}(\tau) \over \eta^{8}(\tau)} + {\vartheta_{10}^{8}(\tau) \over \eta^{8}(\tau)}.
}
and is a singlet under modular transformations.

\ndt The $E_{7}$ characters are $(k=0,1)$:
\eqn\charEseven{
Z^{E7}_{k}(\tau)  = \left({\vartheta_{00}^6(\tau) \over \eta^6(\tau)} + e^{\pi i k} {\vartheta_{01}^6(\tau)\over \eta^6(\tau)}\right) {\vartheta_{k\,0}(2 \tau) \over \eta(\tau)} + {\vartheta_{10}^6(\tau)\over \eta^6(\tau)}  {\vartheta_{k+1\,0}(2 \tau) \over \eta(\tau)}
}
and transform as 
\eqn\esevmod{\eqalign{
\chi_{k}(\tau+1) & = e^{i({3 \over 4} - {k\over 2})\pi} \chi_{k}(\tau) \cr
\chi_{k}(-{1 \over \tau}) & = \sum_{l} {1 \over \rt2} e^{i \pi k l } \chi_{l}(\tau) \cr
}}

\ndt The $E_{6}$ characters are $(k=0,1,2)$:
\eqn\charEsix{
Z^{E6}_{k}(\tau) = {\vartheta^{5}_{00}(\tau) \vartheta_{{2k\over3}0}(3\tau)\over \eta^6(\tau)}  + 
e^{2 \pi i k \over 3}{\vartheta^{5}_{01}(\tau) \vartheta_{{2k\over3}1}(3\tau)\over \eta^6(\tau)} + 
{\vartheta^{5}_{10}(\tau) \vartheta_{1+{2k\over3}\,0}(3\tau)\over \eta^6(\tau)}  
}
and transform as 
\eqn\esevmod{\eqalign{
\chi_{k}(\tau+1) & = e^{i({k^{2} \over 3} - {1\over 2})\pi} \chi_{k}(\tau) \cr
\chi_{k}(-{1 \over \tau}) & = \sum_{l} {1 \over \rt2} e^{i \pi k l } \chi_{l}(\tau) \cr
}}

\listrefs
\end